\documentclass[twocolumn,pra]{revtex4}

\usepackage[latin1]{inputenc}
\usepackage{graphicx}

\begin{document}

\title{Decoherence of matter waves by thermal emission of radiation}
\author{Lucia Hackerm\"{u}ller}
\author{Klaus Hornberger}
\author{Bj\"{o}rn Brezger}
\author{Anton Zeilinger}
\author{Markus Arndt}
\affiliation{Universit\"{a}t Wien, Institut f\"{u}r
Experimentalphysik, Boltzmanngasse 5, A-1090 Wien, Austria}

\date{February 19, 2004.\ Published in: Nature 427, 711--714 (2004)}


\maketitle

{Emergent quantum technologies have led to increasing interest in
decoherence --- the processes that limit the appearance of
quantum effects and turn them into classical phenomena. One
important cause of decoherence is the interaction of a quantum
system with its environment, which 'entangles' the two and
distributes the quantum coherence over so many degrees of freedom
as to render it unobservable. Decoherence theory {[1-4]} has been
complemented by experiments using matter waves coupled to
external photons [5-7] or molecules [8], and by investigations
using coherent photon states [9], trapped ions [10] and electron
interferometers [11,12]. Large molecules are particularly
suitable for the investigation of the quantum-classical
transition because they can store much energy in numerous
internal degrees of freedom; the internal energy can be converted
into thermal radiation and thus induce decoherence. Here we
report matter wave interferometer experiments in which C$_{70}$
molecules lose their quantum behaviour by thermal emission of
radiation. We find good quantitative agreement between our
experimental observations and microscopic decoherence theory.
Decoherence by emission of thermal radiation is a general
mechanism that should be relevant to all macroscopic bodies.}

In this Letter we investigate the decoherence of molecular matter
waves. We change the internal temperature of the molecules in a
controlled way before they enter a near-field interferometer, and
observe the corresponding reduction of the interference contrast.
The idea behind this effort is to demonstrate a most fundamental
decoherence mechanism that we encounter in the macroscopic world:
All large objects, but also molecules of sufficient complexity,
are able to store energy and to interact with their environment
via thermal emission of photons. It is generally believed that
warm macroscopic bodies emit far too many photons to allow the
observation of de Broglie interferences, whereas individual atoms
or molecules can be sufficiently well isolated to exhibit their
quantum nature. However, there must be a transition region
between these two limiting cases. Interestingly, as we show in
this study, C$_{70}$ fullerene molecules have just the right
amount of complexity to exhibit perfect quantum interference in
our experiments [13] at temperatures below 1000 K, and to
gradually lose all their quantum behaviour when the internal
temperature is increased up to 3000 K. We can thus trace the
quantum-to-classical transition in a controlled and quantitative
way. The complexity of large molecules adds a novel quality with
respect to previously performed experiments with atoms [5-7]: the
energy in molecules may be equilibrated in many internal degrees
of freedom during the free flight, and a fraction of the
vibrational energy will eventually be reconverted into emitted
photons. Therefore the internal dynamics of the molecule is also
relevant for the quantum behaviour of the centre-of-mass state.
In contrast to resonance fluorescence, which was investigated
with atoms [5-7], thermal decoherence is omnipresent in
macroscopic systems and it cannot be switched off.

\onecolumngrid

\begin{figure*}[b]
\includegraphics[width=0.72\textwidth]{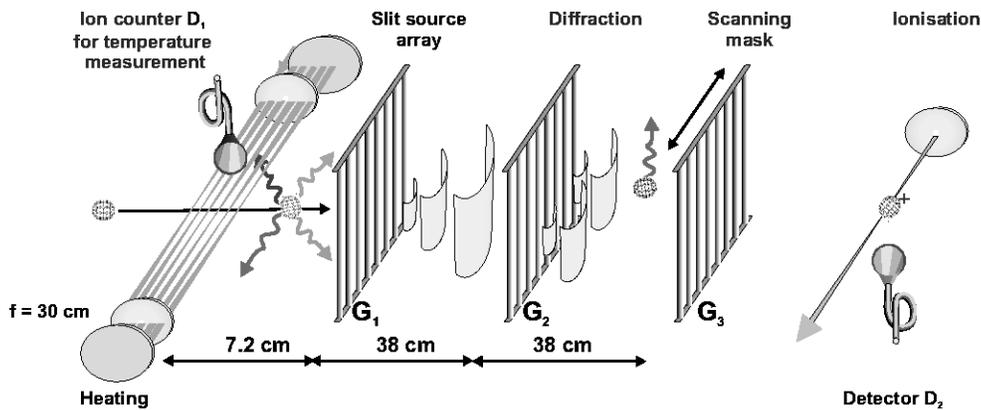}
\caption{Set-up for the observation of thermal decoherence in a
Talbot-Lau molecule interferometer. A fullerene beam passes from
left to right, interacting with a heating stage, a three-grating
(G$_1$--G$_3$) matter-wave interferometer and an ionizing
detection laser beam in D$_2$ (wavelength 488 nm, 1/e$^2$
intensity radius 6.6 $\mu$m, 15 W). The gold gratings have a
period of 990 nm and slit widths of nominally $475\pm20$ nm.
Decoherence of the fullerene matter waves can be induced by
heating the molecules with multiple laser beams (514.5 nm, 40
$\mu$m waist radius, $0-10$ W) before they enter the
interferometer. The resulting molecular temperature can be
assessed by detecting the heating dependent fraction of fullerene
ions using the electron multiplier D$_1$ over the heating stage.}
\end{figure*}

\newpage
\twocolumngrid

The basic set-up of our experiment [14] is sketched in Fig.~1: A
beam of C$_{70}$ molecules is generated by sublimation at about
900 K. The molecules pass a heating stage where they cross a
focused argon ion laser beam up to 16 times. The fullerenes
interact with the laser approximately every 0.3 mm. The laser
heating increases the molecular temperature by 140 K per absorbed
photon. We calculate that they reach up to 5000 K for very short
times, but the re-emission of thermal photons is so efficient
that even the hottest molecules are cooled to below $\sim$ 3000 K
when they enter the interferometer 7.2~cm behind the heating
stage.

The interferometer consists of three identical free-standing gold
gratings with a period of $d=991$ nm. They are separated by the
equal distance of $L=38$ cm, which is the Talbot length $L_{\rm
T}=d^2/\lambda_{\rm dB}$ for a typical de Broglie wavelength of
$\lambda_{\rm dB}=2.6$ pm. The first grating acts as a periodic
array of narrow slit sources, the second one as the diffracting
element, and the third grating is used as a scanning detection
mask, which modulates the molecular density pattern produced by
the Talbot-Lau interference effect [15,16]. The transmitted
molecules are ionized by a blue laser beam (wavelength 488 nm,
6.6 $\mu$m waist), and their intensity $I$ is recorded as a
function of the lateral displacement of the third grating. The
fringe visibility $V=(I_{\rm max}- I_{\rm min})/(I_{\rm
max}+I_{\rm min})$ characterizes the interferogram and thereby
the coherence of the molecular evolution.

\begin{figure}[tb]
\includegraphics[bb=57 33 280 200]{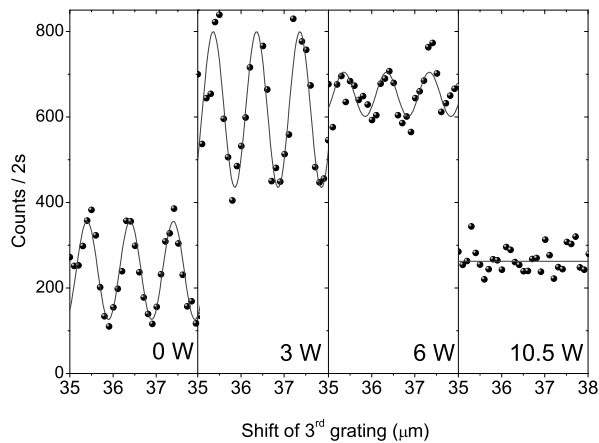}
\caption{Molecule interferograms for C$_{70}$ at 190 m\,s$^{-1}$
for increasing laser heating powers, $P$. The fringe visibility
$V$ decreases with increasing heating power $P$ owing to the
rising emission probability of visible photons: $P$=0\,W
($V$=47\%), $P=3$\,W ($V=29$\%), $P=6$\,W ($V=7$\%), $P=10.5$\,W
(V=0\%). In contrast to that, the absolute count rate grows
initially with increasing $P$. This is due to the fact that the
thermal ionization probability in detector D$_2$ increases with
the temperature of the arriving molecules. At even higher heating
intensities the count rate falls again because of ionization and
fragmentation in the heating stage.}
\end{figure}

The essence of the experiment is to measure the variation of the
interference fringe visibility with heating laser power (Fig.~2).
Two observations can be made: first, the interference contrast
decreases monotonically with increasing power, and vanishes at 10
W. This is the signature of decoherence due to the enhanced
probability for the emission of thermal photons that carry
'which-path' information. Second, we notice that the count rate
also varies considerably. This is explained by the dependence of
the ionization efficiency in the detector D$_2$ on the internal
energy of the fullerenes. It proves that much internal energy
remains in the molecules during their flight through the
apparatus.

In order to confirm quantitatively the interpretations of both
observations, we model the evolution of the distribution of the
internal energies on their way through the apparatus. The
temperature dependence of the spectral photon emission rate
(equation (1) below) then yields the loss of fringe visibility as
predicted by decoherence theory (equation (2) below).

\begin{figure}[tb]
\includegraphics[width=\columnwidth]{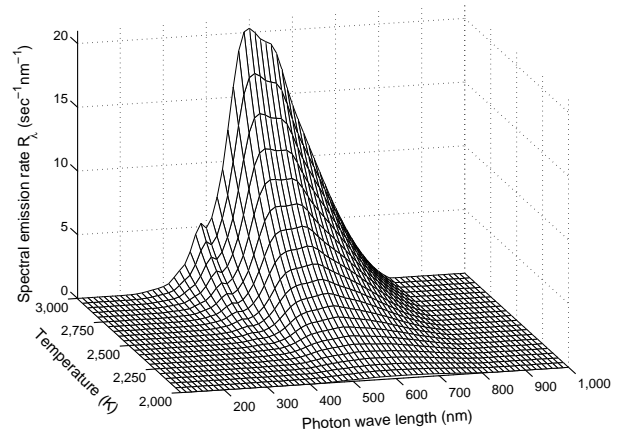}
\caption{Spectral photon emission rate $R_\lambda$ of C$_{70}$
molecules, as used for the calculation of thermal decoherence. We
use the published [25] absorption cross-section for
(S$_0\to$S$_1$) and a heat capacity of $C_V=202 k_{\rm B}$. The
fall-off to short wavelengths is determined by the limited
internal energy of the molecules, while the decrease at long
wavelengths is due to the lack of accessible radiative
transitions at energies below $\sim$1.5 eV. The figure shows that
in the absence of cooling a single molecule at 2500 K travelling
at 190 m s$^{-1}$ (that is, with a transit time of 4 ms through
the interferometer) would emit an integrated number of three
visible photons. This is sufficient to determine the path of the
molecule if the emission occurs close to the second grating.}
\end{figure}

The first photon absorption populates the electronic triplet
state T$_1$ via the excited singlet S$_1$. Given the known
C$_{70}$ triplet lifetimes and non-radiative transition rates
(see ref. 17 and references therein), we can assume that all
further excitation occurs in the triplet system and that the
absorbed excess energy is rapidly transferred to the vibrational
levels. It is known that fullerenes may store more than 100 eV
for a very short time [17], and it was observed that at high
temperatures three different cooling mechanisms start to compete
---  the thermal emission of photons, electrons or C$_{2}$ dimers
[18-22]. These processes are the molecular analogues of the bulk
phenomena known as blackbody radiation, thermionic emission and
evaporative cooling. Following the most recent experimental data
[22], we may safely assume that fragmentation is the least
efficient mechanism. In contrast, thermally activated ionization
is an important mechanism, which we use both in our fullerene
detector [23] and for molecule thermometry, as discussed below.
Nevertheless, we can safely neglect both delayed ionization and
fragmentation for the discussion of the fringe contrast, because
the recoil upon fragmentation and ionization is generally so
large that the affected molecules will miss the narrow detector.
We have also experimentally confirmed that neither C$_{70}^+$
ions nor C$_{68}$ nor smaller fragments from the heating region
are recorded by the detector D$_2$.

However, C$_{70}^+$  ions-and potentially ionized fragments-can
be detected immediately above the heating stage by the electron
multiplier D$_1$ (Fig.~1). To get an estimate of the molecular
temperature distribution, we record the number of ions as a
function of the heating power and of the fullerene velocity. By
comparing the data to a model calculation, we can extract the
parameters that govern the molecular heating of C$_{70}$. Our
model describes the spatial and velocity dependent distribution
of the internal molecular energy by accounting for the stochastic
absorption process, the laser beam characteristics, and the rapid
radiative cooling between the beams as determined by equation (1)
below. It reproduces the detected number of ions in the heating
stage for different laser powers, different numbers of heating
beams and all velocities with the fit parameters for the triplet
absorption cross-section, $\sigma(T_1)=2\times 10^{ -17 }$ cm$^2$,
and the effective Arrhenius constant for ionization, $A_{\rm
ion}=5\times 10^{9}$s$^{ -1}$. The same calculation also describes
the heating dependent increase in count rate at the detector
D$_2$ and thus yields independent information on the temperature
distribution in the molecular beam.

The mean temperature in the beam drops rapidly behind the heating
stage through the emission of thermal photons. The emission of a
continuous photon spectrum has already been observed for
fullerenes in other experiments [18,24]. The equation for the
thermal radiation density differs from the macroscopic Planck law
for several reasons: First, the thermal wavelengths are much
larger than the size of the fullerene, turning it into a coloured
emitter. The mean emission probability is proportional to the
usual mode density factor $\omega^2/(\pi c)^2$ and the known
frequency dependent absorption cross-section [25] $\sigma_{\rm
abs} (\omega)$, assuming that it does not strongly depend on the
internal temperature. Second, the particle is not in thermal
equilibrium with the radiation field. It emits into a cold
environment and stimulated emission does not occur. For this
reason, the statistical factor $1/[\exp(\hbar\omega/k_{\rm
B}T)-1]$ of the Planck formula now would read $\exp(-\hbar\omega/
k_{\rm B}T)$. Third, the 204 vibrational modes of C$_{70}$ do not
constitute an infinite heat bath but have a finite heat capacity
C$_V$. Therefore the emission does not take place at a fixed
temperature, although the internal energy is nonetheless
conveniently characterized by the micro-canonical temperature
$T_{\rm m}$. This leads to a further correction in the spectral
photon emission rate [26], which is now fully described by

\begin{eqnarray}
R_\omega(\omega, T_{\rm m})&=&
\frac{\omega^2}{\pi^2c^2}\,\sigma_{\rm abs}(\omega) \nonumber
\\
&\times&\exp\left[-\frac{\hbar\omega}{k_{\rm B}T_{\rm m}}
-\frac{k_{\rm B}}{2C_V}\left(\frac{\hbar\omega}{k_{\rm B}T_{\rm
m}}\right)^2\right]
\end{eqnarray}
In Fig.~3 we plot the wavelength dependence of $R_\lambda=R_\omega
|{\rm d}\omega/{\rm d}\lambda|$. We observe that at temperatures
below 2000 K the emission rate is negligible, whereas at higher
temperatures the molecules may emit photons whose wavelengths are
comparable to (or even smaller than) the maximum path separation
of $\sim$1 $\mu$m. They transmit (partial) which-path information
to the environment, leading to a reduced observability of the
fullerene wave nature. Around 3000 K the molecules have a high
probability to emit several visible photons yielding sufficient
which-path information to effect a complete loss of fringe
visibility in our interferometer.

\begin{figure}[tb]
\includegraphics[bb=18 76 250 382,width=0.9\columnwidth]{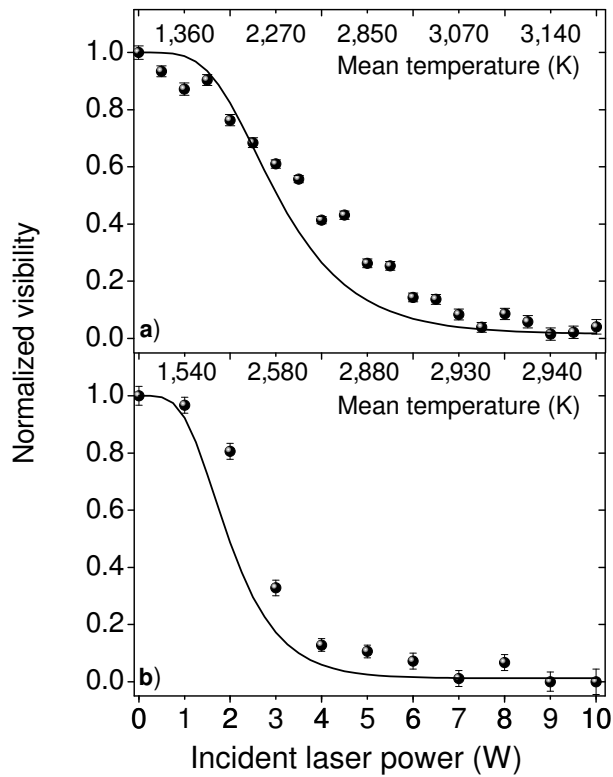}
\caption{Decoherence curves. (a) Interference visibility as a
function of laser heating power (lower scale). The molecular beam
with a mean velocity of v$_{\rm m}$=190 m s$^{-1}$  passes a 50
$\mu$m central height delimiter comparable to the waist
(40~$\mu$m) of the 16 heating laser beams. We observe a rapid
decrease of the fringe visibility with increasing power both in
the experiment (circles) and in theory (solid line). The upper
axis indicates the mean temperature of the molecules when they
enter the interferometer. The maximum contrast without heating
was $V_0$=47\%, which is close to the theoretical value [11]. (b)
Molecules with v$_{\rm m}$=100 m s$^{-1}$, selected by a 150
$\mu$m height delimiter and heated by ten beams of the specified
incident laser power. The qualitative behaviour is the same and
the quantitative agreement with theory is as good as before. The
maximum contrast for this velocity class was $V_0$=19\%. In both
experimental arrangements, a mean number between one and two
visible photons is required to reduce the contrast by a factor of
two.}
\end{figure}

A formal description of this qualitative picture can be given by
decoherence theory. It considers the entanglement of the molecule
with the emitted photon, and shows how coherences vanish once a
trace over the photon state is performed. For objects with
velocity $v$ and temperature evolution $T(t)$ we obtain a
visibility

\begin{eqnarray}
V&=&V_0 \exp\bigg[-\int_0^{2L/v}{\rm d}t\int_0^\infty  {\rm
d}\lambda\, R_\lambda(\lambda,T(t)) \nonumber
\\
&&\quad\quad\quad\times\left\{ 1-{\rm
sinc}\left(2\pi\frac{d}{\lambda}\frac{L-|vt-L|}{L_{\rm T}}\right)
\right\} \bigg],
\end{eqnarray}
as discussed in the Methods section. $V_0$ denotes the
interference contrast in the absence of photon emission. In the
exponential, the sinc function compares the effective molecular
path separation to the radiation wavelength, while the integrals
cover all photon wavelengths $\lambda$ and longitudinal
positions  $vt$  in the interferometer. As a result, the
visibility is reduced exponentially whenever photons are emitted
whose wavelength is sufficiently small to resolve the path
separation. Our predictions for the loss of visibility are
obtained by weighting equation (2) with the previously determined
distribution of temperature evolutions in the molecular beam.

In Fig. 4 we compare our decoherence model with the experiments
by plotting the interference fringe visibility as a function of
the laser power. We observe a strong decrease of the visibility
for molecules at 190 m\,s$^{-1}$, heated by 16 laser beams
(Fig.~4a), and for molecules at 100 m\,s$^{-1}$, heated by 10
laser beams (Fig.~4b).

We also observe good agreement between decoherence theory (solid
line) and the experiment (circles). The experiment is
reproducible within the indicated error bars for a given laser
alignment, but small displacements of the laser focus will
influence the shape and slope of the observed decoherence curve.
The difference between the theoretical and the experimental curve
is of the order of this variation.

In summary, we have presented conclusive empirical and numerical
evidence for observation of the quantum-to-classical transition
of a material object caused by its own emission of thermal
radiation. This auto-localization is a fundamental process
limiting the ultimate observability of quantum effects in
macroscopic objects. However, for nanometre-sized systems
[13,27,28] this mechanism becomes relevant only at high
temperatures, and it is not expected to be a limitation for
interference of objects even considerably larger than the
fullerenes, such as proteins.

\section*{Theoretical Methods}

Equation (2) describes the loss of matter wave coherence due to
the emission of thermal photons. It is obtained by assuming that
the emission is isotropic and that the absorbing walls of the
apparatus are located in the far-field, where the photon position
distribution reflects its momentum distribution. In this case, a
trace over the photon state changes the fullerene centre-of-mass
state $\hat{\rho}$ according to
\begin{equation}
\hat{\rho}\to\hat{\rho}^\prime=\int {\rm d}\mathbf{k}
\frac{p(k)}{4\pi k^2}
\hat{U}_\mathbf{k}\hat{\rho}\hat{U}_\mathbf{k}^\dagger
\end{equation}
where the $\hat{U}_\mathbf{k}=\exp(i\hat{\mathbf{r}}\mathbf{k})$
are momentum translation operators and $p(k)$ is the probability
density for the photon wavenumber $k=2\pi/\lambda$. In the
position representation of the density matrix,
\begin{equation}
\rho'(\mathbf{r}_1,\mathbf{r}_2)\equiv \langle
\mathbf{r}_1|\hat{\rho}'|\mathbf{r}_2\rangle =\langle
\mathbf{r}_1|\hat{\rho}|\mathbf{r}_2\rangle\,\eta(\mathbf{r}_1-\mathbf{r}_2)
\end{equation}
we find from equation (3) that the off-diagonal elements are
reduced by the decoherence function [29]
\begin{equation}
\eta(\mathbf{r}_1-\mathbf{r}_2)=\frac{1}{R_{\rm tot}}
\int_0^\infty {\rm d}\lambda\, R_\lambda(\lambda) \,{\rm
sinc}\left(2\pi\frac{|\mathbf{r}_1-\mathbf{r}_2|}{\lambda}\right).
\end{equation}
Here $p(k)$ is expressed in terms of the spectral emission rate
$R_\lambda$ (see equation (1)) and the total photon emission rate
$R_{\rm tot}$. This sinc-shaped position dependence of $\eta$ is
also found in other experiments with isotropic momentum change
[7,8]. It describes the diffraction limitation of a hypothetical
microscope used to obtain which-path information on the molecules.

In the Talbot-Lau geometry [14,16,27] the final molecular fringe
pattern $w(x)$ is strictly periodic in the grating constant $d$,
and can be expanded as a Fourier series, which reads in the
absence of decoherence
\begin{equation}
w(x)=\sum_\ell C_\ell  \exp(2\pi i\ell x/d).
\end{equation}
Assuming that a single photon emission occurs at the longitudinal
position $z=vt$, a closed expression for the resulting molecular
density pattern can be found. It is obtained by propagating the
molecular density matrix in paraxial approximation first to the
position $z$. We then apply the decohering transformation
(equation (4)) followed by a propagation to the final grating.
For a set-up with equally spaced and identical gratings, the new
fringe pattern is described by a simple modification of the
Fourier coefficients
\begin{equation}
C_\ell\to C_\ell'=C_\ell\,\eta\left(\ell d\frac{L-|vt-L|}{L_{\rm
T}}\right)
\end{equation}

In order to account for more than one photon emission, we make
the Markov assumption that all photon emissions are independent
of each other. Because the modification (equation (7)) is
independent of the molecular density matrix, the change of the
final density pattern is governed by the differential equation
\begin{equation}
\frac{d}{dt} C_\ell=R_{\rm tot}\left[ C_\ell\eta\left(\ell d
\frac{L-|vt-L|}{ L_{\rm T}}\right)-C_\ell \right].
\end{equation}
It describes how the final interferogram is blurred as the time
interval of emission increases. Equation (2) follows then
immediately after taking into account the time dependence of the
emission rate due to cooling (equation (1)) and the fact that for
our grating geometry, with a slit width of 470 nm and grating
constant of 990 nm, only the lowest-order Fourier components
contribute to the fringe visibility [30] $V=2|C_1/C_0|$.

{\small {\bfseries Acknowledgements} { We thank S. Uttenthaler
for his support in an early stage of the experiment. We
acknowledge support by the Austrian START programme, the Austrian
FWF, the European TMR and Marie Curie programmes, and the DFG
Emmy-Noether programme.}

{\bfseries Authors' contributions} { L.H. performed most of the
experiments as a part of her Ph.D. thesis. K.H. developed the
decoherence theory, and made the quantitative comparison between
experiment and theory. }

{\bfseries
Correspondence} {should be addressed to M.A.\\
(markus.arndt@univie.ac.at)} }

\end{document}